\documentclass[aps,prb,preprint,superscriptaddress]{revtex4-1}

\usepackage[dvips]{graphicx}
\usepackage{bm}
\usepackage{braket}
\usepackage{amsmath,amssymb}
\usepackage{comment}
\usepackage{multirow}
\usepackage{threeparttable}
\usepackage[usenames]{color}
\usepackage{ulem}

\def\aa{$\mathrm{\mathring{a}}$}

\begin{document}


\title{Weakly spin-dependent band structures of\\
antiferromagnetic perovskite LaMO$_{3}$ (M $=$ Cr, Mn, Fe)}


\author{Takuya Okugawa}
\affiliation{Department of Physics, Yokohama National University,
79-5 Tokiwadai, Hogogaya-ku, Yokohama 240-8501, Japan}
\author{Kaoru Ohno}
\affiliation{Department of Physics, Yokohama National University,
79-5 Tokiwadai, Hogogaya-ku, Yokohama 240-8501, Japan}
\affiliation{RIKEN Innovation Center, Nakamura Laboratory,
2-1 Hirosawa, Wako, Saitama 351-0198, Japan}
\author{Yusuke Noda}
\affiliation{Center for Materials Research by Information Integration (CMI$^2$),
Research and Services Division of Materials Data and Integrated System (MaDIS),
National Institute for Materials Science (NIMS),
1-2-1 Sengen, Tsukuba, Ibaraki 305-0047, Japan}
\author{Shinichiro Nakamura}
\affiliation{RIKEN Innovation Center, Nakamura Laboratory,
2-1 Hirosawa, Wako, Saitama 351-0198, Japan}


\date{October 16, 2017}

\begin{abstract}
We investigate spin-dependent electronic states of antiferromagnetic (AFM) lanthanum chromite (LaCrO$_{3}$),
lanthanum manganite (LaMnO$_{3}$), and lanthanum ferrite (LaFeO$_{3}$)
using spin-polarized first-principles density functional theory (DFT) with Hubbard  {\sl U} correction.
The band structures are calculated for 15 types of their different AFM structures.
It is verified for these structures that there is a very simple rule to identify
which wave number $\bm{k}$ exhibits spin splitting or degeneracy in the band structure.
This rule uses the symmetry operations that map the up-spin atoms onto the down-spin atoms.
The resulting spin splitting is very small for the most stable spin configuration of the most stable experimental structure.
We discuss a plausible benefit of this characteristic,
{\it i.e.}, the direction-independence of the spin current, in electrode applications.

\vskip20mm

\noindent
This paper was published in Ref. [57]:\\
Takuya Okugawa, Kaoru Ohno, Yusuke Noda, and Shinichiro Nakamura,
``Weakly spin-dependent band structures of antiferromagnetic perovskite LaMO$_3$
(M = Cr, Mn, Fe)'',
Journal of Physics: Condensed Matter {\bf 30}, 075502 (2018).
DOI: 10.1088/1361-648X/aa9e70
\end{abstract}


\maketitle

\section{Introduction}


Materials with perovskite-like structures have many interesting properties
such as ferroelectricity \cite{FER}, superconductivity \cite{SUPER}, photovoltaics \cite{PVC},
catalytic behaviours for the oxygen evolving reaction (OER) or the oxygen reduction reaction (ORR)
\cite{Suntivich1,Suntivich2,Stoerzinger}, spin-dependent transport \cite{SPIN}, and
colossal magnetoresistance (CMR) \cite{CMR,GMR,MVM};
La$_{0.67}$Ca$_{0.33}$MnO$_{x}$ exhibits a huge negative magnetic resistance called
the CMR \cite{CMR,GMR,MVM}, which is applied to magnetic field sensors for reading and writing data in hard disks.  
The formal valence state of La$_{1-x}$Ca$_{x}$MnO$_{3}$ is La$^{3+}$Mn$^{3+}$O$_3^{2-}$ for $x=0$
and Ca$^{2+}$Mn$^{4+}$O$_3^{2-}$ for $x=1$; it becomes a mixed valence state of Mn$^{3+}$ and Mn$^{4+}$ for $x\sim 1/3$,
and, in this case, the ferromagnetic (FM) coupling is favored due to the double exchange mechanism
as explained by Goodenough \cite{Goodenough} and Coey {\it et al}. \cite{MVM}.
Then, just below the Curie temperature, due to the huge magnetic fluctuations, spin current is significantly scattered,
and the electric resistance is high.
However, the applied magnetic field can line up the magnetic orderings, leading to the significant reduction of
the electric resistance.

Not only these complex materials but also its mother material,
lanthanum manganite (I\hspace{-.1em}I\hspace{-.1em}I) (LaMnO$_{3}$), has been investigated in detail,
and it is now well known that LaMnO$_{3}$ has a strongly distorted structure~\cite{MNEX1, MNEX2},
and shows so-called A-type antiferromagnetic (AFM) ordering
below the N\'eel temperature~\cite{MNEX1,MAGN2,MAGN3}.
Similarly, lanthanum chromite (LaCrO$_{3}$) and lanthanum ferrite (LaFeO$_{3}$) have the same structure
as LaMnO$_{3}$~\cite{MAGN3,Okikawa,FECR1,FECR2,FECR3} and they exhibit so-called G-type AFM
ordering below the N\'eel temperature~\cite{MAGN3}.
The OER and ORR of these perovskite materials
have been investigated more recently as an application to electrodes \cite{Suntivich1,Suntivich2,Stoerzinger}.
For these electrochemical reactions, the surface geometry and the surface state related to
Cr$^{2+}$/Cr$^{3+}$/Cr$^{4+}$, Mn$^{2+}$/Mn$^{3+}$/Mn$^{4+}$, and Fe$^{2+}$/Fe$^{3+}$ play an essential role.
So far, many theoretical calculations using density functional theory (DFT)
\cite{Solovyev,Yang,Ravindran,Soltani,LA4F,Gong,Sushiko,Hong,He,Scafetta,Javaid,Wang,Lee,Zhou1,Zhou2,Boateng}
as well as mapping onto tight-binding or spin models \cite{MNMG,MNST2,MNST1}
have been performed for these materials.
Despite these many theoretical studies, there has been only little systematic comparison of all possible structures
of AFM perovskite LaMO$_3$ (M $=$ Cr, Mn, Fe) to find out the similarities and differences among them
in particular in the spin-dependent feature of the electronic states.

There is a naive consensus that, if magnetic ions occupy crystallographically equivalent sites,
the band structures of the up- and down-spin states in antiferromagnetic materials should be the same.
However, is it really true? This question reminds us at least two fundamentally important facts.

One is the fact related to the spontaneous symmetry breaking in infinite systems.
For infinite size antiferromagnetic materials, the magnetic symmetry is broken below the N\'{e}el temperature.
Then, even if all magnetic ions occupy crystallographically equivalent lattice sites,
there appears a clear distinction in the lattice sites of the up- and down-spin ions,
and the positions of their net components are spatially fixed in a alternating way
inside the magnetic domain, although there are of course thermal and quantum fluctuations.
That is, it becomes impossible to reverse all spins in a huge single magnetic domain at once because it requires a huge energy.
This is similar to the situation of a ferromagnetic phase transition;
the only difference is that the lattice is divided into the up-spin A sublattice and the down-spin B sublattice
in an antiferromagnetic case.
Then, the crystal symmetry is lowered to a new symmetry, which is identical to the symmetry of
either the up-spin A sublattice or the down-spin B sublattice;
if the A sublattice symmetry is different from the B sublattice symmetry, the lower symmetry is selected.
In other words, the symmetry coincides with that of a hypothetical crystal having different atomic species,
A and B, respectively, at the A sublattice points and the B sublattice points.
Although X-ray diffraction cannot distinguish these up- and down-spin sites,
neutron diffraction can clearly distinguish them; the pattern of the diffraction spots change
at the N\'{e}el temperature to those with the double spatial periodicity.
However, one should remark that the spontaneous symmetry breaking,
{\it i.e.}, the phase transition never happens for finite size antiferromagnetic materials.
For finite systems, up- and down-spins change in time and no such phase transition takes place.

The other fact is related to the crystal symmetry.
When crystallographically equivalent magnetic up- and down-spin ions are aligned alternately in an antiferromagnetic phase,
their atomic positions can be interchanged, {\it i.e.}, they move onto the other sites by a symmetry operation.
In real space, applying a symmetry operation is equivalent to changing the view point or the view angle.
For example, if the operation is a simple translation, it is clear that the electronic structures
of the up-spin and down-spin electrons are the same everywhere in the Brillouin zone
because they are not affected by the choice of the origin of the real space coordinate system.
However, if the operation $R$ is a two-fold rotation, the electronic structures of the up-spin and down-spin electrons
should be interchanged by the same two-fold rotation $R$ ($= R^{-1}$) in the reciprocal space.
That is, the electronic band structure at the $\bm{k}$ point should be the same as that of the $R\bm{k} (= R^{-1}\bm{k}$) point.
This is the ingredient of our theory.

In either case of the OER and ORR,
the spin current in host materials under a carrier doping, {\it i.e.}, under a
certain shift of the chemical potential, plays an essential role in electrodes.
In our previous study on various AFM MnO$_2$ phases \cite{GR},
we demonstrated that the spin current may have spin-dependent preferable directions
according to the spin-splitting band structures and that,
only for the $\lambda$ phase, all bands are fully degenerate against spin
and no such preferable direction exists in the spin current.
In the $\lambda$ phase, both up- and down-spin currents can coexist in the same direction.
On the other hand, in the $\alpha$, $\beta$, and $\delta$ phases,
there is a spin splitting at the valence band maximum (VBM) and at the conduction band minimum (CBM),
only up- or down-spin can contribute to
the spin current in one direction.
Therefore, fully degenerate bands against spin in AFM phases
must be a favourable factor to enhance the OER and ORR activities;
in fact, in a recent experiment \cite{Robinson}, the OER activity is higher in the $\lambda$ phase than the other MnO$_2$ phases.
In this respect, it is worth investigating
the amount of spin splitting in the band structure
of various AFM perovskite LaMO$_3$ materials also.

Similarly,
it is interesting to consider what happens when the band structure is significantly spin-dependent.
Thanks to de Broglie relationship ($\bm{p}=\hbar\bm{k}$),
the direction dependence of the wave vector $\bm{k}$ is identical to that of the momentum $\bm{p}$.
Then, the property that spin splitting and degeneracy depend on the direction of the wave vector $\bm{k}$
may be utilized for spin-dependent carrier transport or spin selection.
If the VBM or CBM at the wave vector $\bm{k}$ is only composed of up-spin (down-spin) level,
the transport phenomena will be mainly governed by the up-spin (down-spin) carrier with the momentum $\hbar\bm{k}$.
Therefore, spin current has preferable directions in the crystal, and this property may be used for several purposes.
For example, the electrons with different spins could be separated by using the direction-dependent spin current
in electron- or hole-doped materials.  

That is, if the spin-dependance is found in the vicinity of the CBM or VBM, 
one may consider an application of the materials to new spin-control technology or invent a new spin-device.
On the other hand, if the CBM and VBM are both spin degenerate,
the materials are considered to be favourable for the OER or ORR
in a sense that there is no directional preference in the spin current inside the electrodes. 
Therefore, not only from purely academic interest but also from practical interest,
it is quite desirable to investigate
the electronic state of AFM LaMO$_{3}$ in terms of spin-dependent features.

In this paper, we study the spin-dependent
band structures of LaCrO$_{3}$, LaMnO$_{3}$, and LaFeO$_{3}$ using spin-polarized DFT with Hubbard $U$ correction. 
In section II, we survey the possible structures and spin configurations of LaMO$_{3}$.
The computational method adopted in this work is explained in detail in section III.
Sections IV. A and B are devoted, respectively, to the discussion of the resulting optimized structures and magnetic states
and to the general discussion of the spin degeneracy and splitting in energy band diagrams.
Then, sections IV. C, D, and E describe the results of A-type, C-type, and G-type experimental structures.
We explicitly discuss $\bm{k}$-dependent spin splitting and degeneracy for each structure.
The results for less stable structures of LaMO$_3$ are given in Supplementary Data (SD).
In section IV. F, we briefly comment on a preference of the resulting band structures for electrodes.
Finally, section V concludes this paper.

\section{Structures and magnetic configurations of $\bf LaMO_{3}\; (M = Cr, Mn, Fe)$}

\begin{figure}[hbtp!]
\centering
\includegraphics[keepaspectratio, width=70mm, clip]{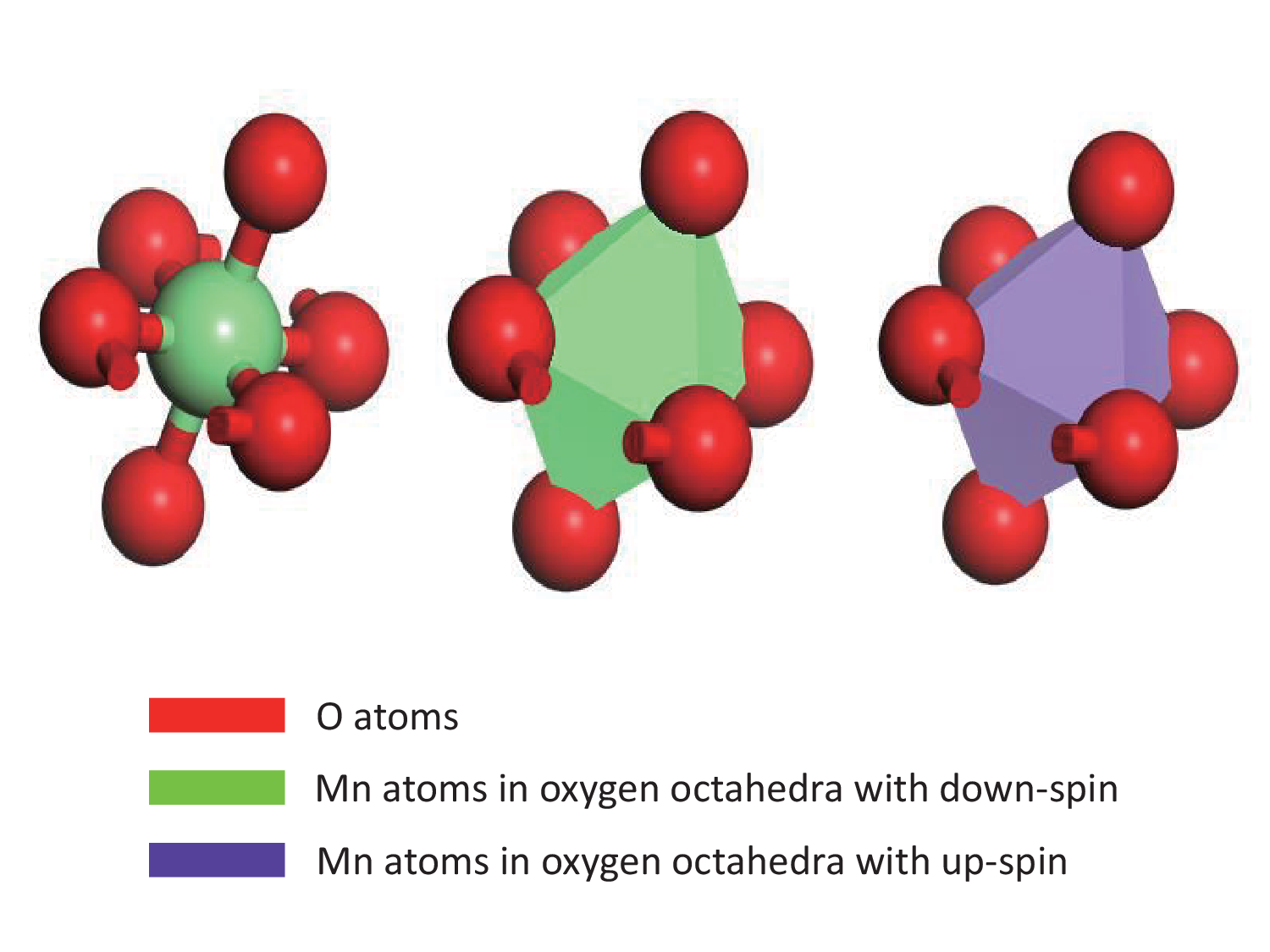}
\caption{Oxygen octahedra surrounding Mn atom.}
\label{fig:OO}
\end{figure}

The strongly distorted structure of LaMO$_{3}$ (with {\sl Pbnm} space group)~\cite{MNEX1, MNEX2}
is derived from the ideal cubic perovskite structure (with {\sl Pm$\bar{3}$m} space group) by deformations
through an intermediate structure that has a pure Jahn-Teller (JT) distortion only (with {\sl P4/mbm} space group)~\cite{MNST1, MNST2}.
For example, in LaMnO$_{3}$, an oxygen octahedron surrounds a Mn atom; see Fig. \ref{fig:OO}.
In what follows, we extensively use this octahedron for the purpose of explaining spin splitting and degeneracy.
In Fig. \ref{fig:OO}, red-coloured balls are O atoms, while green-coloured and magenta-coloured octahedra
(a green-coloured ball in the leftmost figure) denote
Mn atoms with down- and up-spin, respectively.
More details of the structures are explained below. 

\begin{figure}[hbtp!]
\centering
\includegraphics[keepaspectratio, width=90mm, clip]{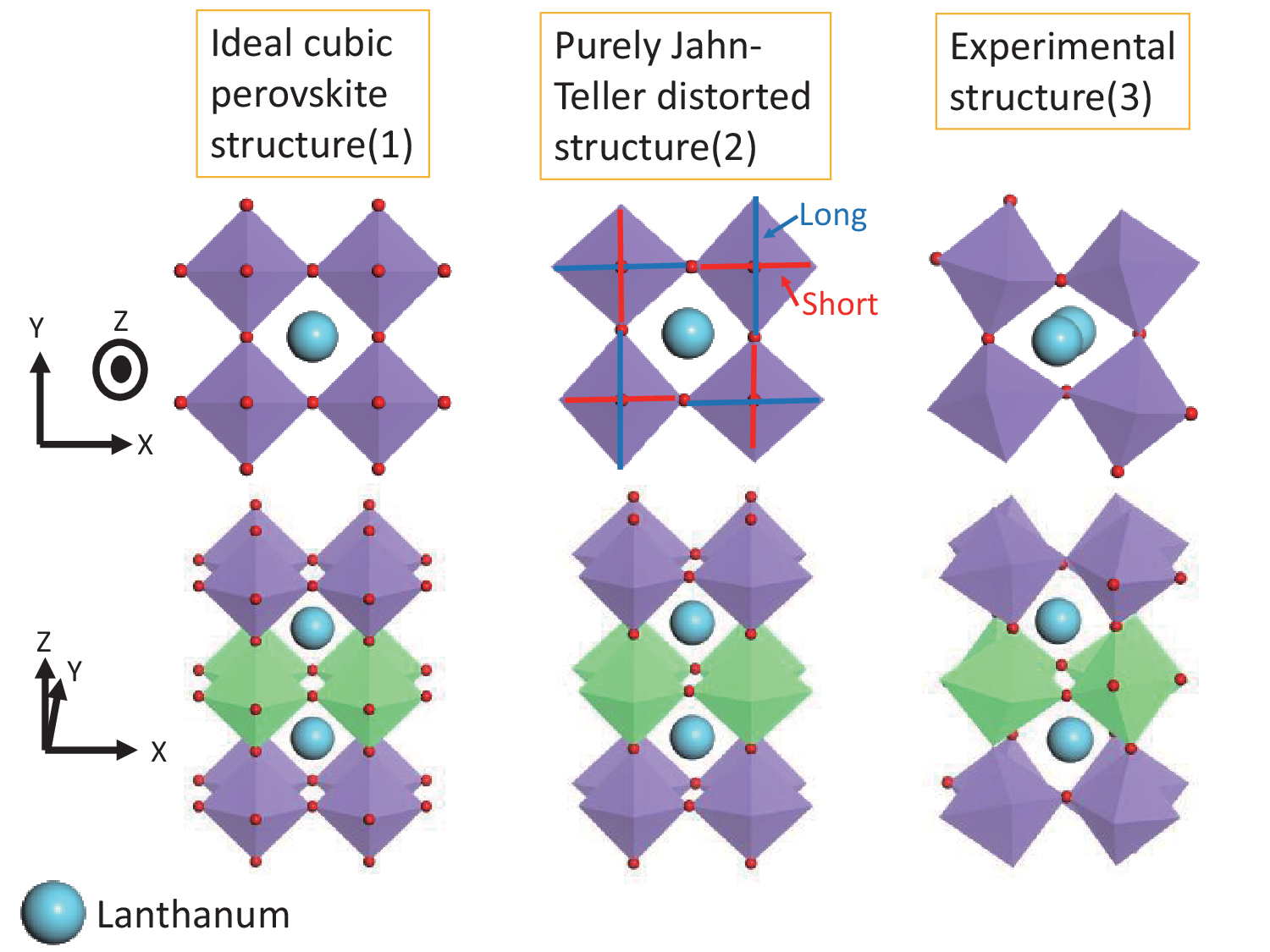}
\caption{Three types of distinct structures studied in this work; see also Supplementary Data (SD).}
\label{fig:3t}
\end{figure}

First of all, three types of the distinct structures~\cite{MNST1, MNST2}
are schematically illustrated in Fig. \ref{fig:3t} and explained as follows: 

\begin{itemize}

\item[(1)] The ideal cubic perovskite structure, in which
all the oxygen octahedra having the same size.
Here we allowed a common elongation or shrink in $z$ direction of oxygen tetrahedra,
so the resulting actual symmetry may become tetragonal (see also SD).

\item[(2)] The purely JT-distorted structure obtained by alternating long and short O-Mn-O distances
in xy-plane from the ideal cubic perovskite structure.
If the O-Mn-O distance along $x$-axis ($y$-axis) in one oxygen octahedron is longer (shorter),
that in adjacent octahedra in the same $xy$-plane is shorter (longer).
Thus, mutually 90 degree rotated (JT-distorted) octahedra are lined up in a checkerboard form.
The resulting symmetry is totally tetragonal.

\item[(3)] The experimental structure realized by rotating oxygen octahedra along $z$-axis and tilting away from $z$-axis.
In addition, lanthanum cations are displaced from their ideal positions within $xy$-plane.
The final symmetry is orthorhombic.

\end{itemize}

\begin{figure}[hbtp!]
\centering
\includegraphics[keepaspectratio, width=90mm, clip]{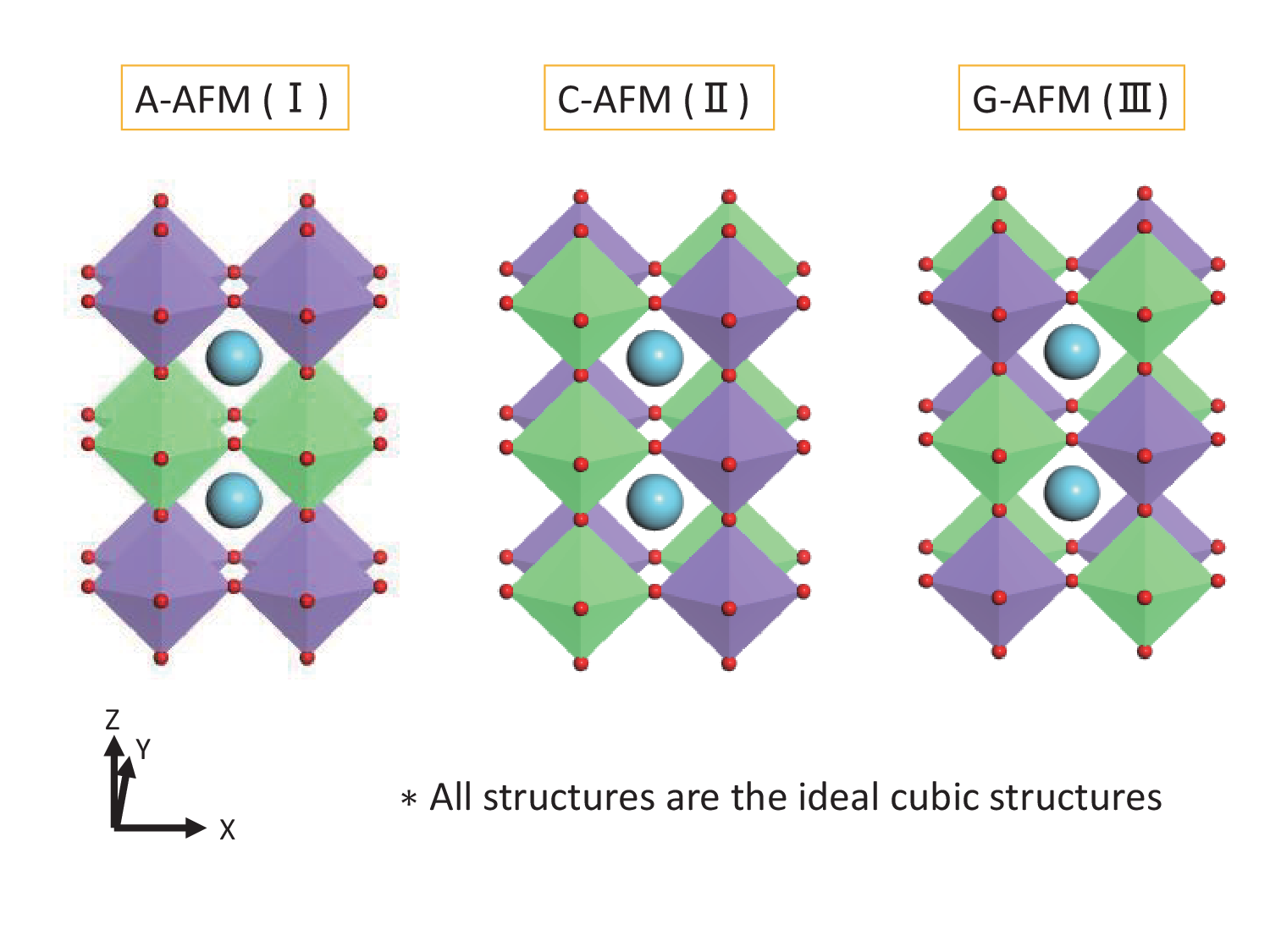}
\caption{Three different magnetic configurations, A-AFM, C-AFM, and G-AFM.}
\label{fig:3m}
\end{figure}

Furthermore, three different magnetic configurations called A-, C- and G-types proposed for LaMnO$_{3}$
by Goodenough \cite{Goodenough} (see also Hotta {\sl et al}.~\cite{MNMG}) are considered in our study also.
Figure \ref{fig:3m} shows these three types of magnetic configuration
in the case of the ideal cubic perovskite structure.

\begin{itemize}

\item[(i)] A-type antiferromagnetic (A-AFM) configuration having a layer-like structure.
Mn atoms inside oxygen octahedra have all up-spin in the first layer,
all down-spin in the second layer, and again all up-spin in the third layer.

\item[(ii)] C-type antiferromagnetic (C-AFM) configuration
having an alternating spin alignments in $xy$-plane.
The spin directions of all Mn atoms are the same along $z$-axis, but in $xy$-plane the neighboring Mn atoms have different spin directions.

\item[(iii)] G-type antiferromagnetic (G-AFM) configuration
having an alternating spin alignment in all directions. The neighboring Mn atoms always have different spin directions.

\end{itemize}
 
Therefore, $3\times3=9$ types of different LaMnO$_{3}$ structures can be considered
and each type of LaMnO$_{3}$ is expected to show distinct electronic states.
In addition, we investigate LaCrO$_{3}$ and LaFeO$_{3}$ in the case of A-, C-, and G-types AFM experimental
(optimized) structures.
Thus, the calculations are carried out for $9+3+3=15$ types of structures. 
All the results except for the experimental structure are presented in Supplementary Data (SD).
Also the results for less stable A- and C-type AFM LaFeO$_{3}$ and LaCrO$_{3}$ are presented in SD.

\section{Computational method}

In this study, we use CASTEP program~\cite{CA1, CA2, CA3} in the Materials Studio package
and perform spin-polarized first-principles DFT calculations
using the on-the-fly generation ultrasoft
pseudopotential approach \cite{Vanderbilt}.
For the exchange-correlation functional,
Perdew-Burke-Ernzerhof (PBE) generalized gradient approximation (GGA)~\cite{GGA}
with Hubbard {\sl U} correction~\cite{HUB, MHUB}
($U=2.5$~eV for Mn, Fe, and Cr and 6.0~eV for La) is used.
Moreover, the calculations with different values of Hubbard {\sl U} parameters (1.5 and 3.5 eV for Mn, 5.0 and 7.0 eV for La)
are carried out in the case of the A-AFM experimental LaMnO$_{3}$ (see Appendix of SD).
Even if any of these {\sl U} parameters are chosen,
our discussion on the spin-dependent band structure does not change. 
Before band calculations, geometrical optimizations were performed together with the unit-cell relaxation.
The cutoff energy for plane-wave basis is set at 571.4~eV for all our calculations.
To get converged results, we set Gaussian smearing at 0.15~eV for geometrical optimization and 0.01~eV for the band calculation.
The initial magnetic moment of M is set at $\pm$5.00$\mu_{B}$. 

For $\bm{k}$-points, we employ Monkhorst-Pack grids~\cite{MONK}
of 6$\times$6$\times$3, 4$\times$4$\times$6, and 4$\times$4$\times$4
for the A-, C-, and G-AFM ideal cubic perovskite structures, respectively.
Also we use $\bm{k}$-point grids of 4$\times$4$\times$3, 3$\times$3$\times$6, and 4$\times$4$\times$6
for the A-, C-, and G-AFM purely Jahn-Teller distorted structures, respectively.
Finally, for all magnetic configurations of the experimental structure,
the 4$\times$4$\times$3 grid is used. 

All these set of parameters are required to get good convergence of the results.
\vspace{0.2in}

\section{Results and Discussion}

\subsection{Optimized structures and magnetic states}

First of all, we summarize the space group and lattice constants obtained by geometrical optimizations
together with the corresponding experimental data (written in parentheses)
for the most stable experimental structure in Table~\ref{struct}.
The estimated total energy of these structures are listed in Table~\ref{TotalEnergy}.
(For the less stable ideal cubic and JT-distorted structures of LaMO$_3$ (M $=$ Cr, Mn, Fe)
as well as the A- and C-AFM experimental structures of LaCrO$_3$ and LaFeO$_3$, see SD.)
\begin{table}[hbtp]
\begin{center}
\caption{Space group and lattice constants ($a,b,c$) for three spin configurations (Spin config.)
of the most stable experimental structure (Exp. struct.).
Values inside parentheses are experimental data \cite{MNEX1,FECR2}.
All other values are calculated by structural optimizations.
}
\begin{threeparttable}
{\tabcolsep = 1.5mm
\begin{tabular}{ccccccc}\hline\hline
Exp.		& Spin 		& Space &&
 \multicolumn{3}{c}{Lattice constants (\AA)} \\ \cline{5-7}
struct.		& config.	& group			&& $a$ 				& $b$ 				& $c$ \\ \hline
			& A-AFM		& $P2_1/c$		&& 5.584 			& 5.616 			& 7.892 \\ \cline{2-7}
LaCrO$_3$	& C-AFM		& $P2_1/m$		&& 5.587 			& 5.594 			& 7.902 \\ \cline{2-7}
$Pbnm$ & \multirow{2}{*}{G-AFM} & \multirow{2}{*}{$P2_1/c$} && 5.595  & 5.573 	& 7.881 \\
			& 			&				&& (5.479)\tnote{a} & (5.513)\tnote{a}	& (7.759)\tnote{a} \\ \hline
	   & \multirow{2}{*}{A-AFM}	& \multirow{2}{*}{$P2_1/c$} && 5.635  & 5.886 	& 7.774 \\
LaMnO$_3$	&			& 				&& (5.532)\tnote{b}	& (5.742)\tnote{b} 	& (7.668)\tnote{b} \\ \cline{2-7}
$Pbnm$		& C-AFM		& $P2_1/m$ 		&& 5.599  			& 6.060 			& 7.735 \\ \cline{2-7}
			& G-AFM		& $P2_1/c$ 		&& 5.596 			& 6.137 			& 7.679 \\ \hline
			& A-AFM		& $P2_1/c$ 		&& 5.631 			& 5.732 			& 7.953 \\ \cline{2-7}
LaFeO$_3$	& C-AFM		& $P2_1/m$		&& 5.617 			& 5.685 			& 7.998 \\ \cline{2-7}
$Pbnm$ & \multirow{2}{*}{G-AFM}	& \multirow{2}{*}{$P2_1/c$} && 5.611  & 5.689 	& 7.970 \\
			&			&				&& (5.553)\tnote{a} & (5.563)\tnote{a} 	& (7.857)\tnote{a} \\
\hline
\end{tabular}
}
\begin{tablenotes}\scriptsize
\item[a] Ref. \cite{FECR2}.
\item[b] Ref. \cite{MNEX1}.
\end{tablenotes}
\end{threeparttable}
\label{struct}
\end{center}
\end{table}
\begin{table}[hbtp]
\begin{center}
\caption{Total energy difference relative to the most stable spin configuration
(A-AFM for LaMnO$_3$ and G-AFM for LaCrO$_3$ and LaFeO$_3$)
of the most stable experimental structure in units of eV.}
\begin{threeparttable}
{\tabcolsep = 2mm
\begin{tabular}{ccccc}\hline\hline
Experimental && \multicolumn{3}{c}{Relative energy (eV)} \\ \cline{3-5}
structure && A-AFM & C-AFM & G-AFM \\ \hline
LaCrO$_3$ && 0.151 
           & 0.069 
           & 0.000 
                           \\
LaMnO$_3$ && 0.000 
           & 0.237 
           & 0.269 
                           \\
LaFeO$_3$ && 0.853 
           & 0.378 
           & 0.000 
                           \\
\hline
\end{tabular}
}
\end{threeparttable}
\label{TotalEnergy}
\end{center}
\end{table}
From this table, it is found that LaCrO$_3$, LaMnO$_3$, and LaFeO$_3$ prefer G-AFM, A-AFM, and G-AFM,
respectively, in accordance with the experimental evidence \cite{MNEX1,MAGN2,MAGN3}.
The calculated band gaps of the most stable experimental structures are given in Table~\ref{gap}
together with the experimental values \cite{Maiti,Mahendiran,Jung,Arima,Bellakki}.

\begin{table}[!h]
\begin{center}
\caption{Energy band gap of the most stable experimental structures.
Values inside parentheses are experimental data \cite{Maiti,Mahendiran,Jung,Arima,Bellakki}.
All other values are calculated results.
}
\begin{threeparttable}
{\tabcolsep = 0.8mm
\begin{tabular}{ccccc}\hline\hline
Experimental &&
 \multicolumn{3}{c}{Band gap (eV)} \\ \cline{3-5}
structure && A-AFM & C-AFM & G-AFM \\ \hline
LaCrO$_3$ && 1.399 & 1.604 & 2.125 (2.8\tnote{a}$\,\,$) \\
LaMnO$_3$ && 0.647 (0.24\tnote{b}, $\,\,$1.2\tnote{c}$\,\,$) & 0.654 & 1.278 \\
LaFeO$_3$ && 0.120 & 0.943 & 1.584 (2.1\tnote{d}, $\,\,$2.4\tnote{e}$\,\,$) \\
\hline
\end{tabular}
}
\begin{tablenotes}\scriptsize
\item[a] Ref. \cite{Maiti}.
\item[b] Ref. \cite{Mahendiran}.
\item[c] Ref. \cite{Jung}.
\item[d] Ref. \cite{Arima}.
\item[e] Ref. \cite{Bellakki}.
\end{tablenotes}
\end{threeparttable}
\label{gap}
\end{center}
\vskip-5mm
\end{table}

If we ignore the magnetic spin configuration, the space-group symmetry
of the experimental structure is $Pbnm$ or equivalently $D_{2h}$, which has 8
symmetry operations including three 180 degree rotations around $x$, $y$, and $z$ axis,
respectively, with translations in [110], [111] and [001] directions:
$T[110]C_2^x$, $T[111]C_2^y$, $T[001]C_2^z$.
However, due to the regular AFM alignment of the electron spins,
the real symmetry is lowered to $P2_1/c$ for A-AFM and G-AFM, and $P2_1/m$ for C-AFM (see Table~\ref{struct}),
which are equivalent to $C_{2v}$ having 4 symmetry operations including only one 180 degree rotation.
Therefore, A-AFM, C-AFM, and G-AFM have, respectively, $T[110]C_2^x$, $T[001]C_2^z$, and $T[111]C_2^y$ (180 degree rotation) only.
In what follows, we will extensively discuss the rest two 180 degree rotations,
{\it i.e.}, $T[111]C_2^y$ and $T[001]C_2^z$ in A-AFM, $T[110]C_2^x$, $T[111]C_2^y$ in C-AFM,
and $T[110]C_2^x$ and $T[001]C_2^z$ in G-AFM,  
which are missing in the lower $P2_1/c$ or $P2_1/m$ symmetry with spin distinction
but existing in the higher $Pbnm$ symmetry without spin distinction.
These two 180 degree rotations in each spin configuration map the up-spin atoms onto the down-spin atoms.

For the most stable structure with the most stable magnetic configuration, Table~\ref{MomentChargeAngle} lists
the spin magnetic moment of M, the Hirshfeld charge \cite{Hirshfeld} of La and M (normalized to give $-2.0$ for O$^{2-}$),
and M-O-M angles; note that there are two different angles M-O1-M (in $z$ direction) and M-O2-M (in $xy$-plane),
because $z$ direction is elongated or shrunk compared to $x$ and $y$ directions.
The $z$ coordinate of O1 (O2) is almost the same as that of La (M).
The spin magnetic moments for M $=$ Cr, Mn, and Fe are very roughly
3$\,\mu_{\rm B}$, 4$\,\mu_{\rm B}$, and 4($\sim$5)$\,\mu_{\rm B}$, respectively,
almost corresponding to the trivalent ionicity (M$^{3+}$), although there is some deviation in Fe.
The trivalent ionicities Cr(III), Mn(III), and Fe(III) are consistent with the calculated results for the Hirshfeld charge;
3.2, 2.4, and 2.7, respectively, for Cr, Mn, and Fe, as shown in Table~\ref{MomentChargeAngle}.
In the same way, we also very roughly identify the valence ionicity of La to be 3 from the Hirshfeld charge;
2.4, 3.1, and 2.9 for La of LaCrO$_3$, LaMnO$_3$, and LaFeO$_3$, respectively.
The M-O1/O2-M angles are all between 90$^{\circ}$ and 180$^{\circ}$,
and would rather closer to 180$^{\circ}$.
According to the theory of Kanamori \cite{Kanamori} on the superexchange interaction
(see also Goodenough \cite{Goodenough}),
two adjacent 3$d$ (M) atoms favour antiparallel (parallel) spin alignment
when the M-O-M angle equals to 180$^{\circ}$ (90$^{\circ}$).
Therefore, the G-type AFM spin configuration might be generally
considered to be more stable than the A-type and C-type spin configurations.
Indeed, the G-type AFM configuration, which is fully AFM in all directions
is realized in LaCrO$_3$ and LaFeO$_3$.
However, for LaMnO$_3$, the A-type AFM configuration, which is AFM in $z$ direction and
FM in $xy$-plane, is energetically more favourable and realized in nature.
Concerning this discrepancy, one may remark the fact that the lattice constant $a$ ($\simeq b$) to $c$ ratio
is close to the ideal value $\sqrt{2}$ for LaCrO$_3$ and LaFeO$_3$
but somewhat smaller than the ideal value ($\sim 1.35$) for LaMnO$_3$ as seen in Table~\ref{struct}.
The larger (shorter) interatomic distance in $xy$-plane (in $z$ direction) keeping the M-O1/O2-M angles,
{\it i.e.} the Jahn-Teller distortion of the octahedra, is the reason of favouring the A-type AFM of LaMnO$_3$.
A possible explanation on this difference is that two $e_g$ levels are empty in Cr$^{3+}$ and fully occupied by
the majority spin in Fe$^{3+}$, but only one of the two $e_g$ levels is occupied by the majority spin in Mn$^{3+}$.
Since the half filling of the $e_g$ state is similar to the mixed valence (Mn$^{3+}$ and Mn$^{4+}$) systems,
there is a double exchange interaction favouring the FM coupling that competes with the superexchange interaction
favouring the AFM coupling, resulting in the A-type AFM due to the Jahn-Teller distortion of LaMnO$_3$
\cite{Cox,Solovyev,LA4F}. 
From Table~\ref{TotalEnergy}, however,
we see that the total energy differences among A-, C-, and G-types of LaMnO$_3$ (LaCrO$_3$)
are less than 0.27~eV (0.16~eV), suggesting that the mechanism favouring these special spin configurations
is related to a very subtle energy balance.
We do not go into further detail because the main purpose of this paper is not to discuss the origin of magnetic interactions
but to discuss the spin degeneracy and splitting of the resulting electronic band structures.


\begin{table*}[hbtp]
\begin{center}
\caption{Calculated and experimental spin magnetic moments of M (M $=$ Cr, Mn, and Fe),
the Hirshfeld charge \cite{Hirshfeld} of La and M (normalized to give $-2.0$ for O$^{2-}$),
and M-O1/O2-M angles, for the most stable structure with the most stable magnetic configuration.
The $z$ coordinates of O1 and O2 are almost the same as those of La and M, respectively.
Available experimental data \cite{MNEX1,MNEX2,MAGN3,Okikawa,FECR3} are also listed for comparison.
}
\begin{threeparttable}
{\tabcolsep = 0.7mm
\begin{tabular}{ccccccccccccc}\hline\hline
Exp. && \multicolumn{2}{c}{Magnetic moment of M ($\mu_{\rm B}$)} &&
 \multicolumn{2}{c}{Hirshfeld charge} &&
 \multicolumn{2}{c}{M-O1-M angle ($^{\circ}$)} && \multicolumn{2}{c}{M-O2-M angle ($^{\circ}$)}
 \\ \cline{3-4} \cline{6-7} \cline{9-10} \cline{12-13}
struct. && calc. & exp. && La & M &&
 calc. & exp. && calc. & exp.
\\ \hline
LaCrO$_3$ && 2.68 & $2.8\pm 0.2$\tnote{a} &&
2.4 & 3.2 &&
157.58 & 161.36\tnote{b} && 159.95 & 161.36\tnote{b} \\
LaMnO$_3$ && 3.67 & $3.9\pm 0.2$\tnote{a}, $\,\,3.7\pm 0.1$\tnote{c} &&
3.1 &2.4 &&
153.15 & $156.0\pm 0.3$\tnote{d} && 154.86 & $156\pm 1$\tnote{d} \\
LaFeO$_3$ && 3.81 & $4.6\pm 0.2$\tnote{a}, $\,\,$3.77\tnote{e} &&
2.9 &2.7 &&
155.44 & 156.8\tnote{e} && 155.36 & 156.8\tnote{e} \\
\hline
\end{tabular}
}
\begin{tablenotes}\scriptsize
\item[a] Ref. \cite{MAGN3}.
\item[b] Ref. \cite{Okikawa}.
\item[c] Ref. \cite{MNEX1}.
\item[d] Ref. \cite{MNEX2}.
\item[e] Ref. \cite{FECR3}.
\end{tablenotes}
\end{threeparttable}
\label{MomentChargeAngle}
\end{center}
\end{table*}

\subsection{Spin degeneracy and splitting}


Here we explain in detail the very simple rule proposed in the preceding paper \cite{GR}
to identify which wave number $\bm{k}$ exhibits spin splitting or degeneracy in the band structure.
The Kohn-Sham (or quasiparticle) equations for the up-spin and down-spin electrons are represented as follows:
\begin{equation}
\label{up-eq}
\left[ \frac{1}{2} {(\bm{k} - i\nabla)}^2 + V_{\uparrow} \right]u_{\bm{k} \uparrow}
  = \epsilon_{\bm{k}\uparrow}u_{\bm{k} \uparrow}, 
\end{equation} 
\begin{equation}
\label{down-eq}
\left[ \frac{1}{2} {(\bm{k} - i\nabla)}^2 + V_{\downarrow} \right]u_{\bm{k} \downarrow}
  = \epsilon_{\bm{k}\downarrow}u_{\bm{k} \downarrow}, 
\end{equation} 
where $\bm{k}$ is the wave vector, $V_{\sigma}$ ($\sigma = \uparrow, \downarrow$) is the effective single-electron potential
including the exchange-correlation potential (or the self-energy),
$u_{\bm{k} \sigma}$ is the periodic part of the eigenfunction,
and $\epsilon_{\bm{k}\sigma}$ is the eigenvalue at one $\bm{k}$-point.

Now, consider symmetry operations $R$ that belong to the (higher symmetry) space group for the atomic geometry ignoring spin orientations
but that do not belong to the (lower symmetry) space group for the spin configuration.
They satisfy
\begin{equation}
RV_{\sigma}R^{-1}= V_{-\sigma},
\label{RVR=V}
\end{equation}
{\it i.e.}, they map the up-spin atoms onto the down-spin atoms and vise versa.
Such symmetry operations, $R$, map the green-coloured
(Mn atom with down-spin) octahedra onto those of magenta-coloured (Mn atom with up-spin);
hereafter we will call such operations the exchange operations.
Then, for the $\bm{k}$-points exhibiting spin degeneracy, one must find at least one such $R$ that satisfies
\begin{equation}
R\bm{k} = \bm{k}.
\label{Rk=k}
\end{equation}
Indeed, $R$ that satisfies both Eqs.~(\ref{RVR=V}) and (\ref{Rk=k}) transforms the square parentheses,
{\it i.e.} the effective Hamiltonian, of Eq.~(\ref{up-eq}) to that of Eq.~(\ref{down-eq}).
Therefore, the eigenvalues of the two equations (\ref{up-eq}) and (\ref{down-eq}) must be the same:
\begin{equation}
\epsilon_{\bm{k}\uparrow} = \epsilon_{\bm{k}\downarrow}.
\end{equation}
That is, spin degeneracy occurs at those $\bm{k}$-points that do not change
({\it i.e.}, satisfies Eq.~(\ref{Rk=k})) under at least one of the exchange operations, $R$. 

On the other hand, if there is no $R$ satisfying Eq.~(\ref{Rk=k}), those $\bm{k}$-points exhibit spin splitting. 
In this case, Eq.~(\ref{Rk=k}) is replaced by
\begin{equation}
R\bm{k} = \bm{k'}.
\label{Rk=k'}
\end{equation} 
Then, $R$ that satisfies both Eqs.~(\ref{RVR=V}) and (\ref{Rk=k'}) transforms the effective Hamiltonian of Eq.~(\ref{up-eq})
at $\bm{k}$ to that of Eq.~(\ref{down-eq}) at $\bm{k'}$.
Therefore, the eigenvalue of up-spin electrons at the $\bm{k}$-point and that of down-spin electrons at the $\bm{k'}$-point
must be the same:
\begin{equation}
\epsilon_{\bm{k}\uparrow} = \epsilon_{\bm{k'}\downarrow}.
\end{equation}
Thus, if Eq.~(\ref{Rk=k'}) holds for one of the exchange operations, $R$, at some $\bm{k}$- and $\bm{k'}$-points,
the up-spin (down-spin) energy at the $\bm{k}$-points and that of the down-spin (up-spin) at the $\bm{k'}$-points become the same.



This rule is a very general one \cite{GR}, and therefore should be applied to all the systems considered here.
In what follows, we discuss spin splitting and degeneracy of the most stable experimental structure in detail.
The results and discussion for less stable cubic and JT-distorted structures of LaMnO$_3$
as well as the A- and C-AFM experimental structures of LaCrO$_3$ and LaFeO$_3$ are given in SD.

\subsection{A-type AFM experimental structure}

\subsubsection{LaMnO$_3$}

\begin{figure*}[hbtp!]
\centering
\includegraphics[keepaspectratio, width=160mm, clip]{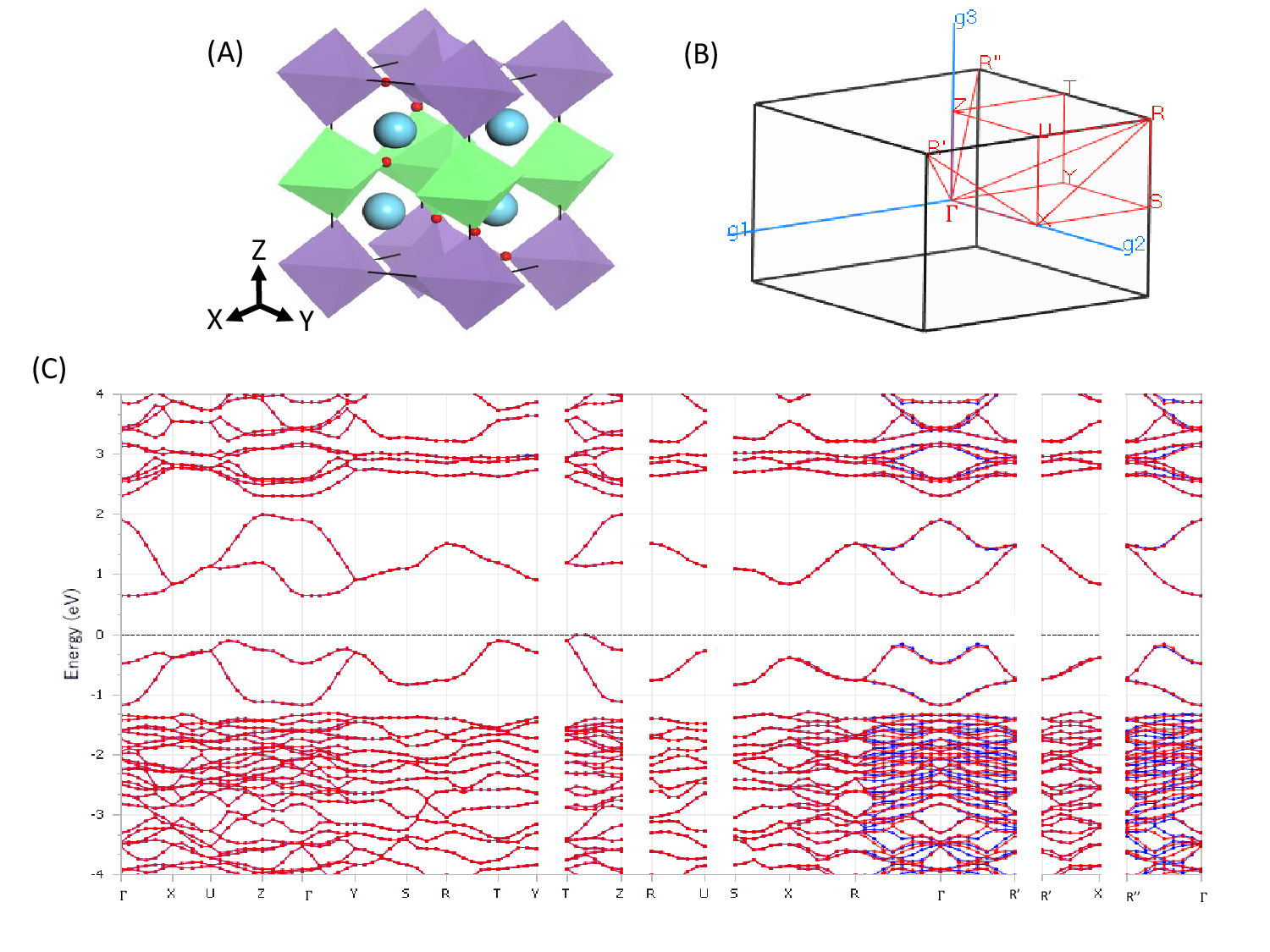}
\caption{(A) Unit cell and (B) Brillouin zone of the A-AFM LaMnO$_3$ experimental ORC structure.
(C) Resulting band structure, in which red and blue lines denote up- and down-spin electron levels, respectively.}
\label{E_A2}
\end{figure*}

The A-AFM experimental structure is the most stable
atomic and spin configuration of LaFeO$_3$, and has a space-group symmetry $Pbnm$
with a simple orthorhombic lattice (ORC, see Fig. \ref{E_A2}(A)) if we ignore the spin configuration.
This space-group symmetry is lowered to  $P2_1/c$ when the spin configuration is taken into account.
This structure with this spin configuration is actually observed experimentally~\cite{MNEX1, MAGN2, MAGN3}.
The lattice constants inferred from our geometrical optimization are $a=5.635\,$\AA, $b=5.886\,$\AA,
$c=7.774\,$\AA{ }(experimental lattice constants~\cite{MNEX1} are $a=5.532\,$\AA, $b=5.742\,$\AA, $c=7.668\,$\AA)
as shown in Table~\ref{struct}.
Figure~\ref{E_A2}(B) is the Brillouin zone of this structure. The band structure is shown in Fig. \ref{E_A2}(C).
There is a band gap of amount 0.647~eV, which should be compared with the experimental values
0.24~eV \cite{Okikawa} and 1.2~eV \cite{MNEX1} as listed in Table~\ref{gap}. 
We added other symmetry $\bm{k}$-lines to those in the conventional Brillouin zone
to discuss the spin-dependent band structure.
Similar to the A- and G-AFM ideal cubic perovskite structures (see SD),
the up- and down-spin energy are completely the same
along $\Gamma$-X-U-Z-$\Gamma$-Y-S-R-T-Y, T-Z, R-U, S-X-R, and R'-X $\bm{k}$-lines.
However, unlike their structures, one can see spin splitting
along R-$\Gamma$-R' and R''-$\Gamma$ $\bm{k}$-lines.
In short, the spin splitting and degeneracy occur depending on $\bm{k}$-points in some manners.
It means that there is a momentum dependence in the band spin splitting and degeneracy.
We confirmed that different sets of Hubbard {\sl U} parameters do not change the $\bm{k}$-points
exhibiting spin degeneracy and those exhibiting spin splitting (see Appendix of SD).
The characteristic of this spin-dependent band structure
can be explained according to the rule described in sections IV. B as follows.


\subsubsection{Spin degeneracy and splitting}

Different from the ideal cubic structure (see SD),
we cannot find any translations as the exchange operation in any direction.
Translations cannot exactly map green-coloured octahedra onto the same structure of magenta-coloured octahedra
because these octahedra are distorted, rotated and tilted. 
However, a combination of 180 degree-rotation $C_{2}$ around $y$-axis and translation along [111] direction
can correctly map green-coloured octahedra onto the same structure of magenta-coloured octahedra:
\begin{equation}
TC_{2}^{y}V_{\downarrow}(TC_{2}^{y})^{-1} = V_{\uparrow}.
\end{equation} 
Under this operation, $\bm{k}$-points on the $\Gamma$-X line in the reciprocal lattice space do not change:
\begin{align}
TC_{2}^{y}\bm{k} = \bm{k} &   & \text{for $\Gamma$-X line.}
\end{align}   
Furthermore, because the Brillouin zone is periodic, $\bm{k}$-points on the Y-S, T-R, and Z-U lines
can go back to the original points
by adding a reciprocal lattice vector $\bm{G}$ if necessary, under the $TC_{2}^{y}$ operation: 
\begin{align}
TC_{2}^{y}\bm{k} = \bm{k}+\bm{G} &   & \text{for Y-S, T-R, and Z-U lines.}
\end{align} 

Moreover, inversion $I$ does not change the structure and the spin configuration;
green-coloured octahedra are mapped onto those of green-coloured by inversion.
Therefore, the combination of 180 degree-rotation around $y$-axis, translation along [111] direction,
and inversion can map green-coloured octahedra onto those of magenta-coloured:
\begin{equation}
\label{ITCY}
ITC_{2}^{y}V_{\downarrow}(ITC_{2}^{y})^{-1} = V_{\uparrow}. 
\end{equation}
Under these operations, $\bm{k}$-points on the $\Gamma$-Y-T-Z-$\Gamma$, X-S-R-U-X, X-R',
and X-R lines in the reciprocal lattice space do not change:
\begin{align}
ITC_{2}^{y}\bm{k} &= \bm{k}            &   & \text{for $\Gamma$-Y-T-Z-$\Gamma$ line,}\\
ITC_{2}^{y}\bm{k} &= \bm{k}+\bm{G} &   & \text{for X-S-R-U-X, X-R', and X-R lines.}
\end{align}

In addition to this combination, we can find a combination of 180 degree-rotation $C_{2}$ around $z$-axis
and translation along [001] direction as the exchange operation.
However, the $\bm{k}$-points on the R-$\Gamma$-R' and $\Gamma$-R'' lines do not map onto themselves
even with this exchange operation.
Thus, we observe spin splitting at such $\bm{k}$-points.
Under the former exchange operation (\ref{ITCY}), $\bm{k}$-points on the $\Gamma$-R'' symmetry line map onto the $\bm{k}$-points
on the $\Gamma$-R symmetry line:
\begin{equation}
ITC_{2}^{y}\bm{k}\;(\Gamma\mbox{-R''})\; = \bm{k}\;(\Gamma\mbox{-R}).
\label{eq14}
\end{equation}
On the other hand, under the latter exchange interaction made by
the combination of the 180 degree-rotation around $z$-axis
and the translation along [001] direction,
those $\bm{k}$-points map onto the $\bm{k}$-points on the $\Gamma$-R' symmetry line:
\begin{equation}
TC_{2}^{z}\bm{k}\;(\Gamma\mbox{-R''})\; = \bm{k}\;(\Gamma\mbox{-R'}).
\end{equation}
Therefore, the up-spin (down-spin) energy on the $\Gamma$-R'' line is the same as the down-spin (up-spin) energy
on the $\Gamma$-R and $\Gamma$-R' lines.

In summary, we find spin degeneracy for all $\bm{k}$-points
on the $\Gamma$-X-U-Z-$\Gamma$-Y-S-R-T-Y, T-Z, R-U, S-X-R, and R'-X lines
and spin splitting on the R-$\Gamma$-R' and R''-$\Gamma$ lines.  

\subsection{C-type AFM experimental structure}

\subsubsection{LaMnO$_3$}

\begin{figure*}[hbtp!]
\centering
\includegraphics[keepaspectratio, width=160mm, clip]{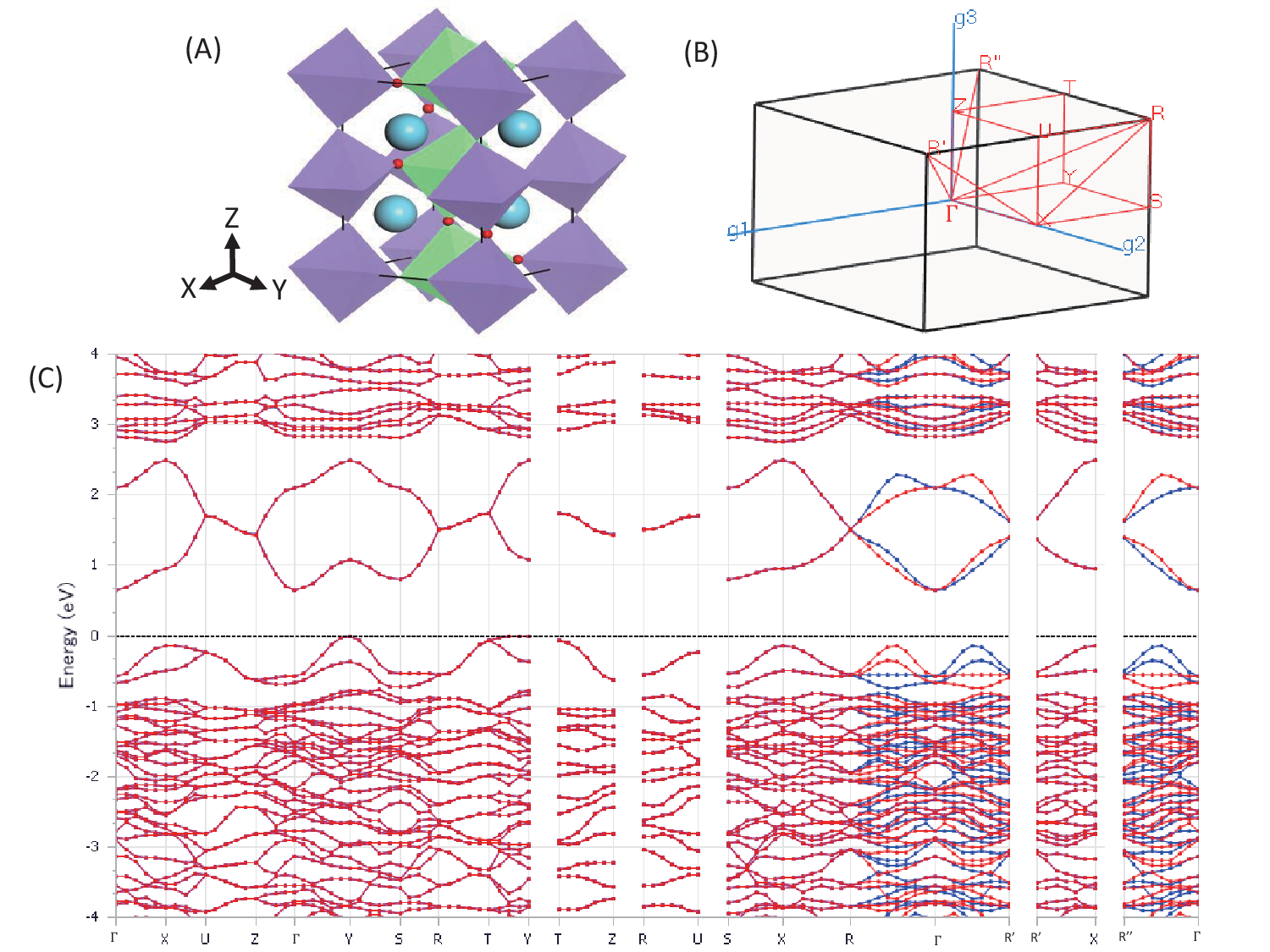}
\caption{(A) Unit cell and (B) Brillouin zone of the C-AFM LaMnO$_3$ experimental ORC structure.
(C) Resulting band structure, in which red and blue lines denote up- and down-spin electron levels, respectively.}
\label{E_C2}
\end{figure*}

The C-AFM LaMnO$_3$ experimental structure has a space-group symmetry $Pbnm$
with a simple orthorhombic lattice (ORC, see Fig. \ref{E_C2}(A)),
which is lowered to $P2_1/m$ by taking into account the spin configuration.
The relaxed lattice constants are $a=5.599\,$\AA, $b=6.060\,$\AA, $c=7.735\,$\AA{ }as
shown in Table~\ref{struct}.
Figure~\ref{E_C2}(B) is the Brillouin zone of this structure.
The band structure is shown in Fig. \ref{E_C2}(C).
The amount of the band gap is 0.654~eV as in Table~\ref{gap}.
Similar to the A-AFM LaMnO$_3$ experimental structure,
spin splitting occurs along R-$\Gamma$-R' and R''-$\Gamma$ symmetry lines
and spin degeneracy is seen in the other $\bm{k}$-point region.

\subsubsection{Spin degeneracy and splitting}

This type of the structure has (A) a combination of 180 degree-rotation $C_{2}$
around $y$-axis and translation along [111] direction
and (B) a combination of 180 degree-rotation $C_{2}$ around $x$-axis
and translation along [110] direction as the exchange operations.
The discussion for the former combination (A) is the same as in the A-AFM LaMnO$_3$ experimental structure,
and we find spin degeneracy on the $\Gamma$-X-U-Z-$\Gamma$-Y-S-R-T-Y, T-Z, R-U, S-X-R, and R'-X lines.
On the other hand, the latter exchange operation (B), which does not exist in the A-AFM LaMnO$_3$ experimental structure,
even cannot map the $\bm{k}$-points on the R-$\Gamma$-R' and $\Gamma$-R'' lines onto themselves.
Therefore, the $\bm{k}$-points on the R-$\Gamma$-R' and R''-$\Gamma$ lines exhibit spin splitting.


In this case, the discussion about the $\bm{k}$-points on the $\Gamma$-R'' and $\Gamma$-R lines
is the same as the case of the A-AFM experimental structure
because the same exchange operation (\ref{ITCY}) and the relationship (\ref{eq14}) exist.
Regarding the $\bm{k}$-points on the $\Gamma$-R and $\Gamma$-R' lines,
a combination of 180 degree-rotation $C_{2}$ around $x$-axis,
translation along [110] direction, and inversion, {\it i.e.}, a combination of (B) and inversion, can be used:
\begin{equation}
ITC_{2}^{x}\bm{k}\;(\Gamma\mbox{-R})\; = \bm{k}\;(\Gamma\mbox{-R'}).
\end{equation}
(Note that inversion itself is the symmetry operation in the space group.)
Thus, the up-spin and down-spin energies on the $\Gamma$-R line exchange with those on the $\Gamma$-R'' and $\Gamma$-R' lines.  

\subsection{G-type AFM experimental structure}

\subsubsection{LaMnO$_3$}

\begin{figure*}[hbtp!]
\centering
\includegraphics[keepaspectratio, width=160mm, clip]{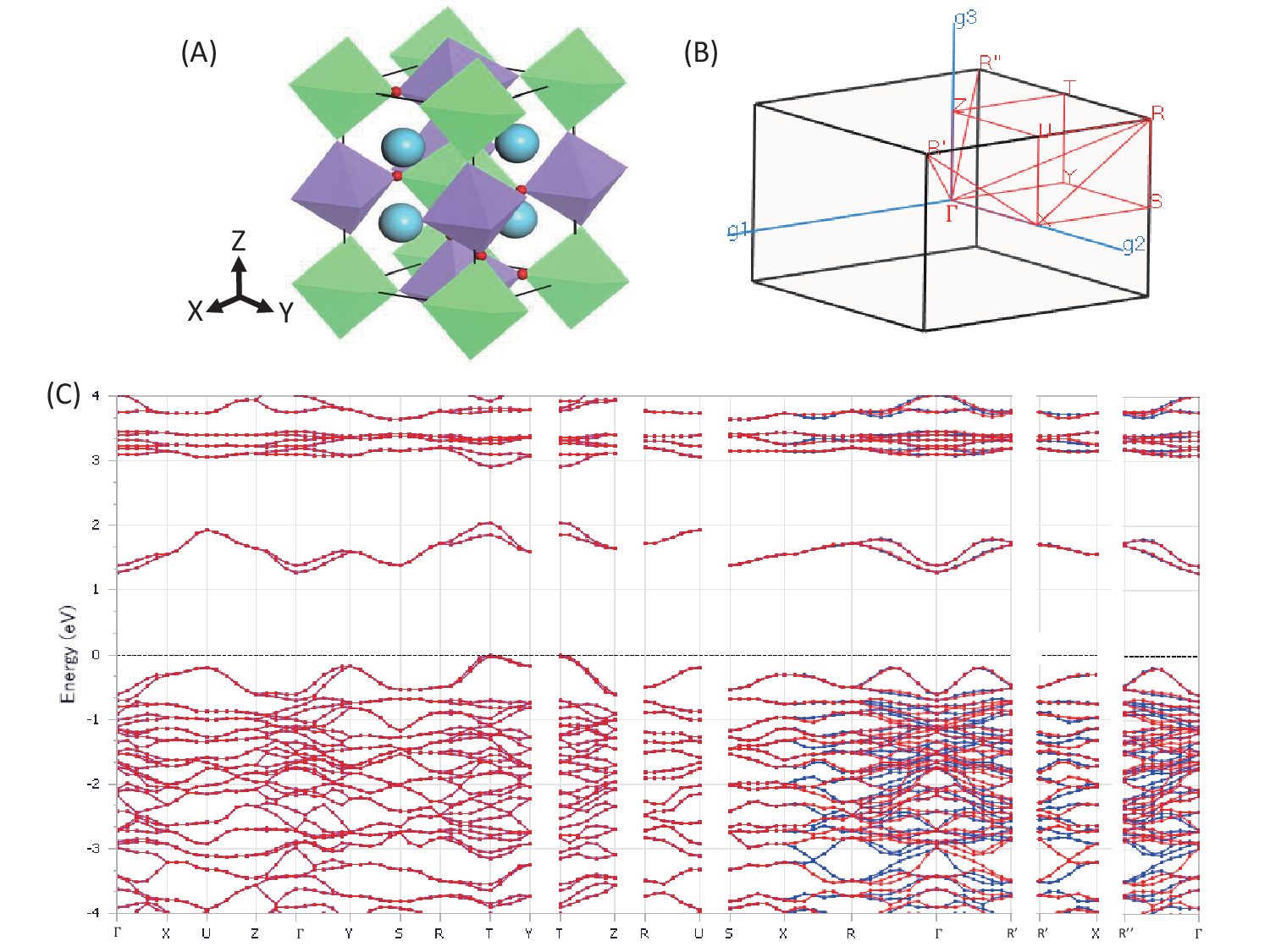}
\caption{(A) Unit cell and (B) Brillouin zone of the G-AFM LaMnO$_3$ experimental ORC structure.
(C) Resulting band structure, in which red and blue lines denote up- and down-spin electron levels, respectively.}
\label{E_G2}
\end{figure*}

The G-AFM LaMnO$_3$ experimental structure has a space-group symmetry $Pbnm$
with a simple orthorhombic lattice (ORC, see Fig. \ref{E_G2}(A)),
which is lowered to $P2_1/c$ by taking into account the spin configuration.
The relaxed lattice constants are
$a=5.596\,$\AA, $b=6.137\,$\AA, $c=7.679\,$\AA{ }as shown in Table~\ref{struct}.
Figure~\ref{E_G2}(B) is the Brillouin zone of this structure.
The band structure is shown in Fig. \ref{E_G2}(C).
There is a band gap of amount 1.278~eV as written in Table~\ref{gap}.
While the up- and down-spin energy levels are completely the same
along the $\Gamma$-X-U-Z-$\Gamma$-Y-S-R-T-Y, T-Z, R-U, and S-X $\bm{k}$-lines,
they are different along the X-R-$\Gamma$-R', R'-X, and R''-$\Gamma$ lines. 

\subsubsection{Spin degeneracy and splitting}

The exchange operations for this type of structure are (A) a combination of 180 degree-rotation $C_{2}$
around $x$-axis and translation along [110] direction,
and (B) a combination of 180 degree-rotation $C_{2}$
around $z$-axis and translation along [001] direction. 
By the combination (A),
\begin{equation}
TC_{2}^{x}V_{\downarrow}(TC_{2}^{x})^{-1} = V_{\uparrow}.
\end{equation}
Unchanged $\bm{k}$-points are on the $\Gamma$-X, Y-S, T-R, and Z-U lines:
\begin{align}
TC_{2}^{x}\bm{k} &= \bm{k} & & \text{for the $\Gamma$-X line,}\\
TC_{2}^{x}\bm{k} &= \bm{k}+\bm{G} & & \text{for the Y-S, T-R, and Z-U lines.} 
\end{align}
Due to the inversion symmetry, operation of $I$ does not change the geometrical structure and the spin configuration.
Therefore we can consider
\begin{align}
\label{ITCX}
ITC_{2}^{x}V_{\downarrow}(ITC_{2}^{x})^{-1} = V_{\uparrow},
\end{align}
and, under this operation, $\bm{k}$-points on $\Gamma$-X-U-Z-$\Gamma$ and Y-S-R-T-Y lines do not change:
\begin{align}
ITC_{2}^{x}\bm{k} &= \bm{k} & & \text{for the $\Gamma$-X-U-Z-$\Gamma$ line,}\\
ITC_{2}^{x}\bm{k} &= \bm{k}+\bm{G} & & \text{for the Y-S-R-T-Y line.}
\end{align}
Even if we use the combination (B), it changes $\bm{k}$-points on the X-R-$\Gamma$-R'-X and R''-$\Gamma$ lines.
Therefore, we see spin degeneracy for the $\bm{k}$-points
on the $\Gamma$-X-U-Z-$\Gamma$-Y-S-R-T-Y, T-Z, R-U, and S-X-R lines and spin splitting
on the X-R-$\Gamma$-R'-X and R''-$\Gamma$ lines.   


By using the exchange operation (\ref{ITCX}), we can obtain the relationship
\begin{align}
ITC_{2}^{x}\bm{k}\;(\Gamma\mbox{-R})\;& = \bm{k}\;(\Gamma\mbox{-R'}),\\
ITC_{2}^{z}\bm{k}\;(\Gamma\mbox{-R'})\; & = \bm{k}\;(\Gamma\mbox{-R''}).
\end{align}
Accordingly, the up-spin (down-spin) energy on the $\Gamma$-R' is the same as the down-spin (up-spin) energy
on the $\Gamma$-R and $\Gamma$-R''.

\subsubsection{LaCrO$_3$}

\begin{figure*}[hbtp!]
\centering
\includegraphics[keepaspectratio, width=160mm, clip]{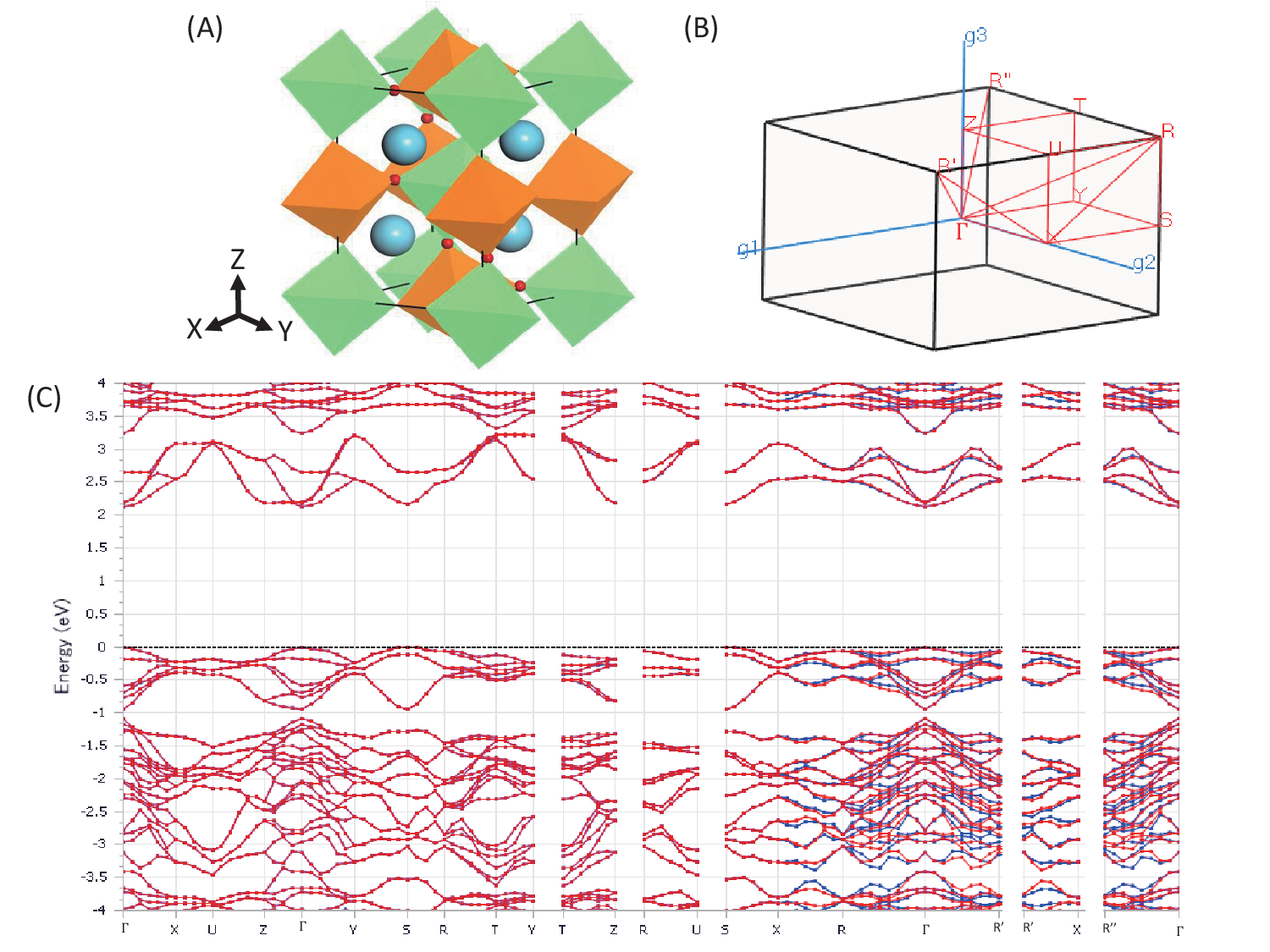}
\caption{(A) Unit cell of the G-AFM LaCrO$_3$ experimental ORC structure.
Ocher- and green-coloured octahedra denote Cr atoms with up- and down-spin, respectively
(the definition is also applied to SD).
(B) Corresponding Brillouin zone and (C) resulting band structure,
in which red and blue lines denote up- and down-spin electron levels, respectively.}
\label{Cr_G2}
\end{figure*}

The G-AFM LaCrO$_3$ experimental structure has a space-group symmetry $Pbnm$
with a simple orthorhombic lattice (ORC, see Fig. \ref{Cr_G2}(A)),
which is lowered to $P2_1/c$ when the spin configuration is taken into account.
Actually, this structure with the spin configuration is obtained experimentally~\cite{FECR2}.
After the geometrical optimization, the lattice constants are $a=5.595\,$\AA, $b=5.573\,$\AA,
$c=7.881\,$\AA{ }(experimental lattice constants are $a=5.479\,$\AA, $b=5.513\,$\AA, $c=7.759\,$\AA)
as shown in Table~\ref{struct}.
The Brillouin zone of this structure is shown in Figure~\ref{Cr_G2}(B), and
the resulting band structure is shown in Fig.~\ref{Cr_G2}(C).
According to our calculation, the band gap becomes 2.125~eV, which should be compared with the experimental values
2.8~eV \cite{MAGN3}; see Table~\ref{gap}.

The assignment of the spin splitting and degeneracy at each $\bm{k}$-point
is exactly the same as the G-AFM experimental structure of LaMnO$_3$.
Therefore, we see spin degeneracy for the $\bm{k}$-points
on the $\Gamma$-X-U-Z-$\Gamma$-Y-S-R-T-Y, T-Z, R-U, and S-X-R lines and spin splitting
on the X-R-$\Gamma$-R'-X and R''-$\Gamma$ lines.
For the results of A-type and C-type experimental structures of LaCrO$_3$, see SD.

\subsubsection{LaFeO$_3$}

\begin{figure*}[hbtp!]
\centering
\includegraphics[keepaspectratio, width=160mm, clip]{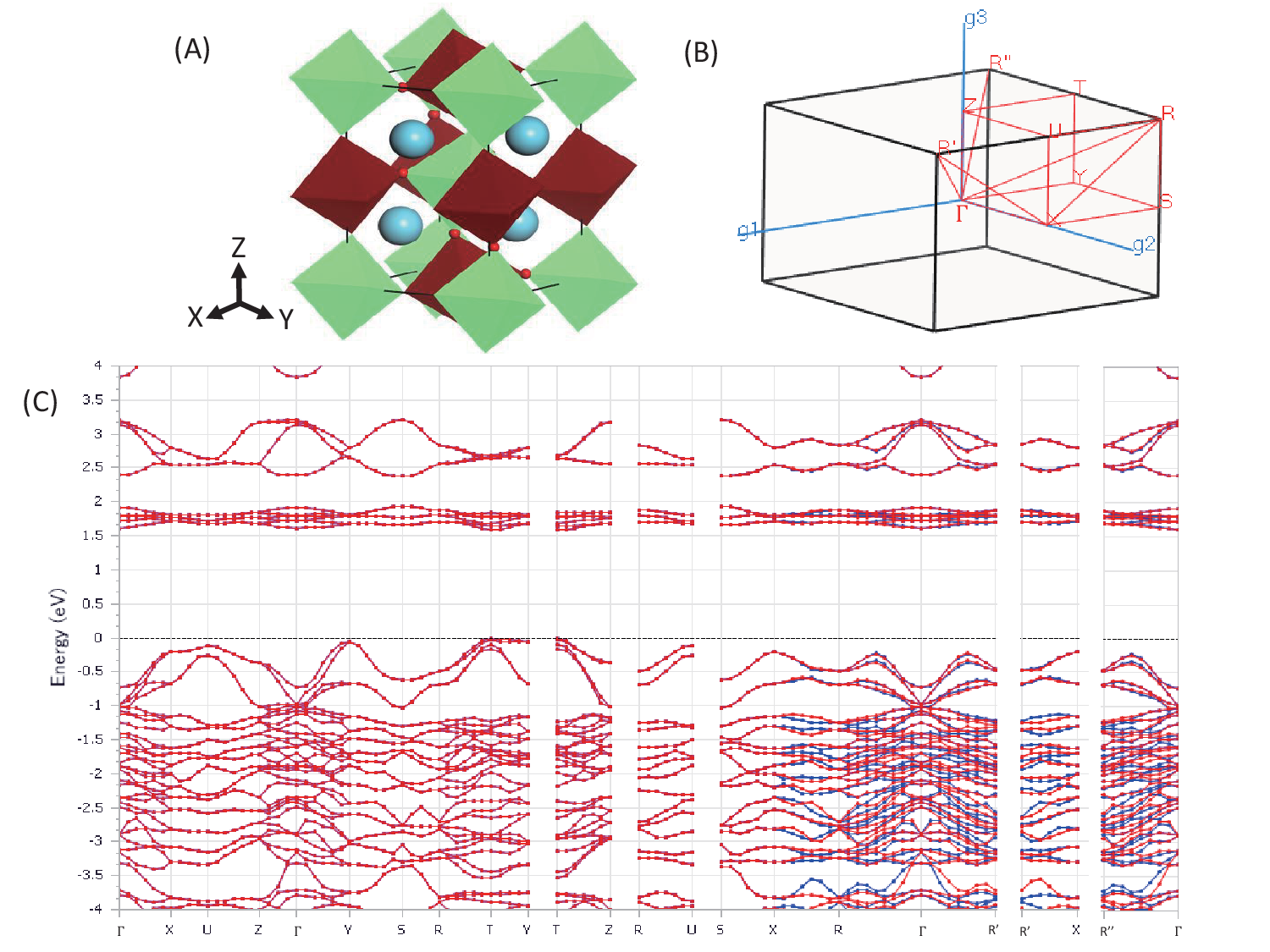}
\caption{(A) Unit cell of the G-AFM LaFeO$_3$ experimental ORC structure.
Brown- and green-coloured octahedra denote Fe atoms with up- and down-spin, respectively
(the definition is also applied to SD).
(B) Corresponding Brillouin zone and (C) resulting band structure,
in which red and blue lines denote up- and down-spin electron levels, respectively.}
\label{Fe_G2}
\end{figure*}

The G-AFM LaFeO$_3$ experimental structure has a space-group symmetry $Pbnm$
with a simple orthorhombic lattice (ORC, see Fig. \ref{Fe_G2}(A)),
which is lowered to $P2_1/c$ when the spin configuration is taken into account.
The spin configuration and the structure are the same as those reported experimentally~\cite{FECR2, FECR3}.
The lattice constants obtained by the geometrical optimization are $a=5.611\,$\AA, $b=5.689\,$\AA,
$c=7.970\,$\AA{ }(experimental lattice constants are $a=5.553\,$\AA, $b=5.563\,$\AA, $c=7.857\,$\AA)
as shown in Table~\ref{struct}.
The Brillouin zone and the resulting band structure are shown in Figures~\ref{Fe_G2}(B) and (C), respectively.
The estimated band gap is 1.584~eV, which should be compared with the experimental values
2.1~eV \cite{MNEX2} and 1.2~eV \cite{FECR3}; see Table~\ref{gap}.

The spin splitting and degenerate $\bm{k}$-points are exactly the same as the G-AFM experimental structures 
of LaMnO$_3$ and LaCrO$_3$.
Therefore, we see spin degeneracy for the $\bm{k}$-points
on the $\Gamma$-X-U-Z-$\Gamma$-Y-S-R-T-Y, T-Z, R-U, and S-X-R lines and spin splitting
on the X-R-$\Gamma$-R'-X and R''-$\Gamma$ lines.
For the results of the A-type and C-type experimental structures of LaFeO$_3$, see SD.

\subsection{A brief comment on a preference for electrodes}

We found that the resulting band structures of the most stable atomic and spin configurations
of LaMO$_3$ (M $=$ Cr, Mn, Fe) are very similar to each other and almost spin degenerate everywhere in the Brillouin zone,
although a very small spin splitting appears in some $\bm{k}$-points.
Therefore, after a carrier doping in these LaMO$_3$ materials,
we expect that there is no preferable direction in the spin current
and both up- and down-spin can equally contribute to the spin current in every direction.
In order to discuss the OER and ORR activities of these materials,
it is essential to consider the spin-dependent chemical reactions at surfaces
as have been recently investigated by several authors \cite{Suntivich1,Suntivich2,Stoerzinger,Wang,Lee}.
This is because every oxygen molecule associated with these reactions is spin triplet.
Due to the local AFM ordering, spin directions are alternatively aligned also at surfaces.
Therefore, to accelerate these reactions at the electrode surfaces, it would be also essential to 
evenly supply the spin-independent current in every direction in the bulk.
In this sense, the almost spin-independent band structure in the bulk can be considered
as a factor to enhance the OER and ORR reactivity of these materials. 
Indeed, these materials have recently attracted great interest in the OER and ORR applications \cite{Suntivich1,Suntivich2,Stoerzinger}.
On the other hand, we should say that these materials are unfortunately not suitable
for a device to separate the up-spin and down-spin currents in different directions,
which could be possible if there was a significant spin dependence in the band structure.

\section{Conclusion}

We investigated the spin-dependent band structures for 15 types of antiferromagnetic (AFM) LaCrO$_3$, LaMnO$_3$,
and LaFeO$_3$ by spin-polarized DFT with GGA+{\sl U} calculations.
Geometry optimizations and band calculations were carried out to the 15
AFM LaCrO$_3$, LaMnO$_3$, and LaFeO$_3$ structures.

According to our calculation of the total energy, G-type magnetic configuration is the most stable
for both LaFeO$_3$ and LaCrO$_3$, which is consistent with the experimental results \cite{MAGN3}.
In the case of LaMnO$_3$, A-type magnetic configuration is favored in
the total energy comparison, which is also consistent with the experimental results \cite{MNEX1,MAGN2,MAGN3}.
The results for the less stable cubic and JT-distorted structures of LaMnO$_3$
as well as the A- and C-AFM experimental structures of LaCrO$_3$ and LaFeO$_3$ are given in SD.

From the resulting band structures,
there is a very interesting difference between spin degeneracy and splitting
depending on $\bm{k}$-points ({\it i.e.} electron or hole momentum) in the Brillouin zone.   
In the case of the A- and G-AFM LaMnO$_3$ ideal cubic perovskite and purely Jahn-Teller distorted structures,
there is spin degeneracy in the whole Brillouin zone (see SD).
In contrast, spin splitting can be seen in all $\bm{k}$-lines and $\bm{k}$-points
for the C-AFM LaMnO$_3$ ideal cubic perovskite structure (see SD).
Furthermore, there are both spin degenerate and splitting $\bm{k}$-points in the other LaMO$_3$ (M=Mn, Fe, and Cr) structures.   

The difference between the spin splitting and degenerate $\bm{k}$-points in the present systems
is also successfully explained by the very simple rule we found in the preceding paper \cite{GR}:
If a space-group symmetry operation $R$ can map the effective (Kohn-Sham or more sophisticated self-energy-related) potential
of an up-spin (down-spin) electron onto that of a down-spin (up-spin) electron, $RV_{\uparrow}R^{-1}=V_{\downarrow}$
$(RV_{\downarrow}R^{-1}=V_{\uparrow})$, and a certain wave vector $\bm{k}$ is not changed under this operation, {\it i.e.}
$R\bm{k}=\bm{k}$, then spin degeneracy can be seen at that $\bm{k}$-point in the band structure.
Otherwise, for two different wave vectors $\bm{k}$ and $\bm{k'}$ which satisfy $R\bm{k}=\bm{k'}$ under the relationship
$RV_{\sigma}R^{-1}=V_{-\sigma}$, the up-spin band energy at the $\bm{k}$-point
and that of down-spin at $\bm{k'}$-point are the same, and {\sl vice versa}.
In the preceding paper \cite{GR}, this rule was confirmed for the four phases of MnO$_{2}$ only.
Now, we have succeeded in confirming the validity of this rule for variety
of magnetic and structural types of AFM LaMnO$_3$, LaFeO$_3$, and LaCrO$_3$. 

In the present study, we found that the band structures of the most stable atomic and spin configurations
of LaMO$_3$ (M $=$ Cr, Mn, Fe) are very similar to each other and almost spin degenerate everywhere in the Brillouin zone,
although a very small spin splitting appears in some $\bm{k}$-points.
Therefore, after a carrier doping in these LaMO$_3$ materials,
we expect that there is no preference in the direction in the spin current
and both up- and down-spins can equally contribute in every direction.
This fact may favour the OER and ORR reactions of these electrodes,
because they can evenly supply the spin-independent current in every direction.
On the other hand, we should say that these materials are not suitable
for a device to separate the up-spin and down-spin currents in different directions,
which could be possible if there were a significant spin dependence in the band structure.

\section*{Acknowledgements}
We are very grateful to Dr. Abhijit Chatterjee and Dr. Riichi Kuwahara at Dassault Syst\`emes BIOVIA
for a lot of helpful discussions and support.
This work was also supported in part by the Ministry of Education, Culture, Sports, Science and Technology (MEXT) Japan
as a social and scientific priority issue (Creation of new functional devices and high-performance
materials to support next-generation industries) to be tackled by using post-K computer
for the use of the supercomputer facilities of the Supercomputer Centers
at the Institute for Solid State Physics, the University of Tokyo,
at Hokkaido University, and at IMR, Tohoku University
(Project IDs. hp160072, hp160234, hp170268, and hp170190).

\vskip5mm

\noindent
{\bf Supplementary Data (SD):} Supplementary Data are available from ``Public Full-text''
at ResearchGate Web page:\\
\url{https://www.researchgate.net/publication/321416556}


\vskip5mm


\begin{thebibliography}{99}
\bibitem{FER} S. Mathews, R. Ramesh,  T. Venkatesan, and J. Benedetto,
Science, $\bm{276}$, 238 (1997).

\bibitem{SUPER} M. K. Wu, J. R. Ashburn, C. J. Torng, P. H. Hor, R. L. Meng, L. Gao, Z. J. Huang, Y. Q. Wang, and C. W. Chu,
Phys. Rev. Lett. {\bf 58}, 908 
(1987).

\bibitem{PVC} Z. Guo, Y. Wan, M. Yang, J. Snaider, K. Zhu, L. Huang,
Science {\bf 356}, 59 
(2017).

\bibitem{Suntivich1} J. Suntivich, K. J. May, H. A. Gasteiger, J. B. Goodenough and Y. Shao-Horn,
Science {\bf 334}, 1383 
(2011).

\bibitem{Suntivich2} J. Suntivich, H. A. Gasteiger, N. Yabuuchi, H. Nakanishi, J. B. Goodenough and Y. Shao-Horn,
Nat. Chem. {\bf 3}, 546 
(2011).

\bibitem{Stoerzinger} K. A. Stoerzinger, M. Risch, J. Suntivich, W. M. Lu, J. Zhou, M. D. Biegalski, H. M. Christen, Ariando,
T. Venkatesan and Y. Shao-Horn,
Energy Environ. Sci. {\bf 6}, 1582 
(2013).


\bibitem{SPIN} S. A. Wolf, D. D. Awschalom, R. A. Buhrman, J. M. Daughton, S. von Molnar, M. L. Roukes, A. Y. Chtchelkanova, D. M. Treger, Science $\bm{294}$, 1488 (2001).

\bibitem{CMR} S. Jin, T. H. Tiefel, M. McCormack, R. A. Fastnacht, R. Ramesh, and L. H. Chen, Science $\bm{264}$, 413 (1994).

\bibitem{GMR} M. N. Baibich., J. M. Broto, A. Fert, F. Nguyen Van Dau, F. Petroff, P. Eitenne, G. Creuzet, A. Friederich, and J. Chazelas, Phys. Rev. Lett. $\bm{61}$, 2472 (1988).


\bibitem{MVM}J. M. D. Coey, M. Viret, S. von Molnar, Advances in Physics, $\bm{48}$, 167 (1999).

\bibitem{Goodenough} J. B. Goodenough, Phys. Rev. {\bf 100}, 564 (1955).

\bibitem{MNEX1} J. B. A. A. Elemans, B. Van Laar, K. R. Van der Veen, and B. O. Loopstra, J. Solid State Chem. $\bm{3}$, 238 (1971).

\bibitem{MNEX2} P. Norby, I. K. Andersen, E. K. Andersen, and N. Andersen, J. Solid State Chem. $\bm{119}$, 191 (1995).

\bibitem{MAGN2} E. O. Wollan and W. C. Koehler, Phys. Rev. $\bm{100}$, 545 (1955).

\bibitem{MAGN3} W. C. Koehler and E. O. Wollan, J. Phys. Chem. Solids $\bm{2}$, 100 (1957).


\bibitem{Okikawa} K. Oikawa, T. Kamiyama, T. Hashimoto, Y. Shimojyo, and Y. Morii,
J.  Solid State Chem. {\bf 154}, 524 
(2000).




\bibitem{FECR1}S. Geller and E. A. Wood, Acta Cryst. $\bm{9}$, 563 (1956).

\bibitem{FECR2}R. Dogra, A. C. Junqueira, R. N. Saxena, A. W. Carbonari, J. Mestnik-Filho, and M. Moralles,
Phys. Rev. B $\bm{63}$, 224104 (2001).

\bibitem{FECR3}G. L. Beausoleil II, P. Price, D. Thomsen, A. Punnoose, R. Ubic, S. Misture, and D. P. Butt,
J. Am. Ceram. $\bm{97}$, 228 (2014).


\bibitem{Solovyev} I. Solovyev, N. Hamada, and K. Terakura,
Phys. Rev. Lett. {\bf 76}, 4825 
(1996).

\bibitem{Yang} Z.-Q. Yang, Z. Huang, L. Ye, and X. Xie,
Phys. Rev. B {\bf 60}, 15674 
(1999).

\bibitem{Ravindran} P. Ravindran, R. Vidya, H. Fjellv{\aa}g, A. Kjekshus,
J. Crys. Growth {\bf 268}, 554 
(2004).

\bibitem{Soltani} N. Soltani, S. M. Hosseini, and A. Kompany,
Physica B {\bf 404}, 4007 
(2009).

\bibitem{LA4F} T. Hashimoto, S. Ishibashi, and K. Terakura,
Phys. Rev. B {\bf 82}, 045124 
(2010).

\bibitem{Gong} S. Gong and B.-G. Liu,
Phys. Lett. A {\bf 375}, 1477 
(2011).

\bibitem{Sushiko} P. V. Sushko, L. Qiao, M. Bowden, T. Varga, G. J. Exarhos, F. K. Urban III,
D. Barton, and S. A. Chambers,
Phys. Rev. Lett. {\bf 110}, 077401 
(2013).

\bibitem{Hong} J. Hong, A. Stroppa, J. \'{I}$\tilde{\rm n}$iguez, S. Picozzi, and D. Vanderbilt,
Phys. Rev. B {\bf 85}, 054417 
(2012).

\bibitem{He} J. He and C. Franchini,
Phys. Rev. B {\bf 86}, 235117 
(2012).

\bibitem{Scafetta} M. D. Scafetta, A. M. Cordi, J. M. Rondinelli, and S. J. May,
J. Phys.: Condens. Matter {\bf 26}, 505502 
(2014).

\bibitem{Javaid} S. Javaid and M. J. Akhtar,
J. Appl. Phys. {\bf 116}, 023704 
(2014).

\bibitem{Wang} Y. Wang and H.-P. Cheng,
J. Phys. Chem. C {\bf 117}, 2106 
(2013).

\bibitem{Lee} Y.-L. Lee, M. J. Gadre, Y. Shao-Horn, and D. Morgan,
Phys. Chem. Chem. Phys. {\bf 17}, 21643 
(2015).

\bibitem{Zhou1} Y.-J. Zhou, Z. L\"{u}, B. Wei, Z.-B. Zhu, Q.-Y. Xie, Y.-Q. Li, and W.-H. Su,
Fuel Cells {\bf 13}, 1040 
(2013).

\bibitem{Zhou2} Y.-J. Zhou, Z. L\"{u}, B. Wei, S. Xu, D. Xu, and Z. Yang,
Ionics {\bf 22}, 1153 
(2016).

\bibitem{Boateng} I. W. Boateng, R. Tia, E. Adei, N. Y. Dzade, C. R. A. Catlow, and N. H. de Leeuw,
Phys. Chem. Chem. Phys. {\bf 19}, 7399 
(2017).

\bibitem{MNMG} T. Hotta, S. Yunoki, M. Nayr, and E. Dagotto,
Phys. Rev. B {\bf 60}, R15009 
(1999).

\bibitem{MNST2} C. Ederer, C. Lin, and A. J. Millis,
Structural distortions and model Hamiltonian parameters: From LSDA to a tight-binding description of LaMnO3
Phys. Rev. B {\bf 76}, 155105 
(2007).

\bibitem{MNST1} R. Kov\'{a}\v{c}ik and C. Ederer,
Phys. Rev. B {\bf 81}, 245108 
(2010).

\bibitem{GR} Y. Noda, K. Ohno, S. Nakamura, Phys. Chem. Chem. Phys. $\bm{18}$, 13294 (2016).

\bibitem{Robinson} D. M. Robinson, Y. B. Go, M. Mui, G. Gardner, Z. Zhang, D. Mastrogiovanni, E. Garfunkel,
J. Li, M. Greenblatt, and G. C. Dismukes,
J. Am. Chem. Soc. {\bf 135}, 3494 
(2013).




\bibitem{CA1} M. D. Segall, P. J. D. Lindan,
M. J. Probert, C. J. Pickard, P. J. Hasnip,
S. J. Clark, and M. C. Payne,
J. Phys. Cond. Matter. $\bm{14}$, 2717 (2002).

\bibitem{CA2} 
CASTEP in Materials Studio, Biovia, Dassault Syst\`{e}mes.

\bibitem{CA3} S. J. Clark, M. D. Segall, C. J. Pickard, P. J. Hasnip, M. J. Probert, K. Refson, and M. C. Payne,
Z. Kristallographie $\bm{220}$, 567 (2005). 



\bibitem{Vanderbilt} D. Vanderbilt, Phys. Rev. B {\bf 41}, 7892 (1990).

\bibitem{GGA} J. P. Perdew, K. Burke and M. Ernzerhof, Phys. Rev. Lett., $\bm{77}$, 3865 (1996).

\bibitem{HUB} V. I. Anisimov, J. Zaanen, O. K. Andersen, Phys. Rev. B $\bm{44}$, 943 (1991). 
    
\bibitem{MHUB} V. I. Anisimov, J. Zaanen, Phys. Rev. B $\bm{52}$, 5467 (1995). 

\bibitem{MONK} H. J. Monkhorst and J. D. Pack, Phys. Rev. B $\bm{13}$, 5188 (1976).


\bibitem{Maiti} K. Maiti and D. D. Sarma, Phys. Rev. B {\bf 54}, 7816 (1996).

\bibitem{Mahendiran} R. Mahendiran, A. K. Raychaudhuri, A. Chainani, D. D. Sarma,
and S. B. Roy, Appl. Phys. Lett. {\bf 66}, 233 (1995).

\bibitem{Jung} J. H. Jung, K. H. Kim, D. J. Eom, and T. W. Noh, E. J. Choi, J. Yu, Y. S. Kwon, and Y. Chung,
Phys. Rev. B {\bf 55}, 15489 
(1997).

\bibitem{Arima} T. Arima, Y. Tokura, and J. B. Torrance, Phys. Rev. B {\bf 48}, 17006 (1993).

\bibitem{Bellakki} M. B. Bellakki, B. J. Kelly, and V. Manivannan, J. Alloys Compd. {\bf 489}, 64 (2010).


\bibitem{Hirshfeld} F. L. Hirshfeld,
Theoret. Chim. Acta (Berl.) {\bf 44}, 129- 
(1977).

\bibitem{Kanamori} J. Kanamori, J. Phys. Chem. Solids {\bf 10}, 87 (1959).

\bibitem{Cox} P. A. Cox, {\sl Transition Metal Oxides: An Introduction to Their
Electronic Structure and Properties}, Clarendon Press, Oxford, 1995.

\bibitem{Okugawa} T. Okugawa, K. Ohno, Y. Noda, and S. Nakamura,
J. Phys.: Condens. Matter {\bf 30}, 075502 (2018).

\end{thebibliography}
\end{document}